\documentclass[aps,prl,reprint,superscriptaddress]{revtex4-1}

\usepackage{graphicx}
\usepackage{float}
\usepackage{siunitx}
\usepackage{array,booktabs}

\begin{document}


\title{Experimental Measurements of Ion Heating in Collisional Plasma Shocks and Interpenetrating 
Supersonic Plasma Flows}


\author{Samuel J. Langendorf}
\email[]{samuel.langendorf@lanl.gov}
\affiliation{Los Alamos National Laboratory, Los Alamos, NM 87545}

\author{Kevin C. Yates}
\affiliation{University of New Mexico, Albuquerque, NM 87131}
\affiliation{Los Alamos National Laboratory, Los Alamos, NM 87545}

\author{Scott C. Hsu}
\email[]{scotthsu@lanl.gov}
\affiliation{Los Alamos National Laboratory, Los Alamos, NM 87545}

\author{Carsten Thoma}
\affiliation{Voss Scientific, Albuquerque, NM 87108}

\author{Mark Gilmore}
\affiliation{University of New Mexico, Albuquerque, NM 87131}


\date{\today}

\begin{abstract}

We present time-resolved measurements of ion heating due to collisional plasma shocks and 
interpenetrating 
supersonic plasma flows, which are formed by the oblique merging of two coaxial-gun-formed plasma 
jets. Our study was repeated using four jet species:  N, Ar, Kr, and Xe.  In conditions with 
small interpenetration between jets, the observed peak ion temperature $T_i$ is consistent with the 
predictions of collisional plasma-shock theory, showing a substantial elevation of $T_i$ above the 
electron temperature $T_e$ and also the subsequent decrease of $T_i$ on the classical ion--electron 
temperature-equilibration time scale. In conditions of significant interpenetration between jets, such that 
shocks  do not apparently form, the observed peak $T_i$ is still appreciable and greater than $T_e$,
but much lower than that predicted by collisional plasma-shock theory.  Experimental results are compared
with multi-fluid plasma simulations.

\end{abstract}


\maketitle


Shocks are a fundamental feature of supersonic plasma flows and affect
the energy balance and dynamical evolution of physical systems in which the shocks are 
embedded, e.g., in astrophysical systems \cite{blandford1987particle, bell1978acceleration, 
ryu2003cosmological, markevitch2002textbook} or in high-energy-density (HED)
\cite{drake06} and inertial-confinement-fusion (ICF) \cite{atzeni04} experiments.
Differing in two key respects from hydrodynamic shocks, plasma
shocks (1)~are mediated either by classical Coulomb collisions between plasma particles (collisional
plasma shock \cite{jukes57,jaffrin64}) or by collective effects such as the Weibel instability 
\cite{weibel59,fox2013filamentation,huntington15} (collisionless plasma shock \cite{tidman71}),
and (2)~are more complex due to
the coupled interactions of electrons, ions (sometimes multiple species), 
electromagnetic fields, and radiative and equation-of-state (EOS) effects. 
This Letter focuses on ion heating in
unmagnetized collisional plasma shocks and interpenetrating supersonic
plasma flows, where radiative/thermal losses and EOS effects are important.  
Related recent experiments include colliding 
railgun plasma jets \cite{merritt2013experimental,merritt14,moser2015experimental}, 
wire-array Z pinches \cite{harvey-thompson12,swadling2013oblique,swadling14prl}, and 
laser ablation of solid targets 
\cite{rinderknecht18prl}. The latter is also used to study collisionless shocks 
\cite{romagnani08,kuramitsu11,ryutov12,ross2012characterizing,ross17}.  
The study of interpenetrating, colliding plasma flows has a long history, e.g., 
\cite{rumsby74,pollaine92,wan97}. Time-resolved ion-temperature data were not reported in nor
were the focus of the prior works.

This Letter presents the first detailed diagnostic study of the time evolution
of ion temperature $T_i$ and ion heating due to 
unmagnetized collisional plasma shocks and interpenetrating supersonic plasma flows,
with sufficient detail to compare with theory and simulation across species and collisionality regimes.
These new fundamental data 
are valuable for validating and improving first-principles
modeling of these phenomena, e.g., \cite{casanova1991kinetic,berger91,vidal93}, which are crucial
for advancements in modeling HED/ICF experiments and a range of
astrophysical plasmas.
There are significant disagreements among different codes and models \cite{keenan17pre,keenan18pop},
possibly due to specific choices of collisionality, transport, and EOS models and/or their 
implementations.  Although HED/ICF experiments have different absolute plasma parameters, our 
experiments are in a similar regime with respect to collisionality and EOS, such that the same models
and codes are applicable.

\begin{figure}[!tb]
\includegraphics[width=3.25truein]{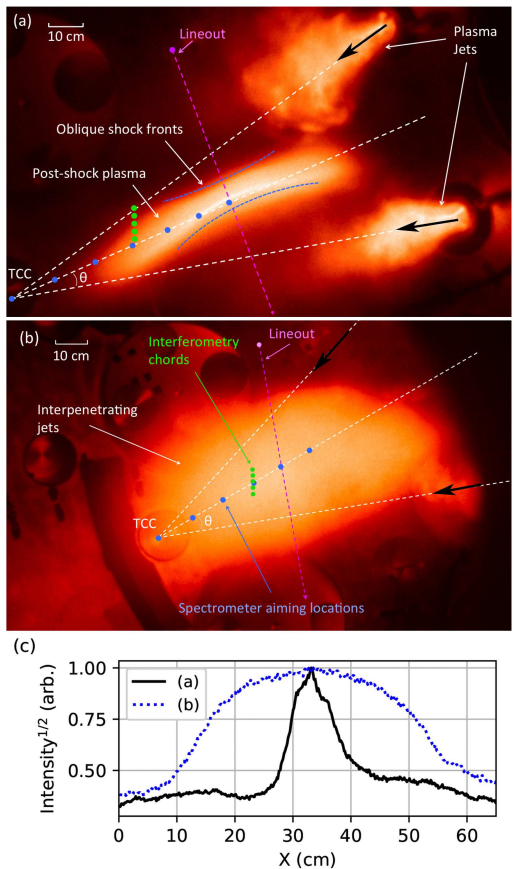}
\caption{Fast-camera, visible-light images (10-ns exposure, log intensity, false color) of two merging
Ar plasma jets (black arrows indicate direction of travel)
with (a)~$\theta=11.6^{\circ}$ (shot 2559, $t=38$~$\mu$s), showing the formation of 
oblique, collisional plasma shocks, and (b)~$\theta=20.5^\circ$ (shot 
1570, $t=36$~$\mu$s), showing Ar jet--jet interpenetration (large region of diffuse emission)  
without apparent shock formation. 
Diagnostic chord positions (green and blue dots) and target-chamber center 
(TCC) are shown. (c)~Lineouts of the square root of intensity correspond to the magenta dotted lines in 
(a) and (b).}
\label{fig:setup}
\end{figure}

Results presented here were obtained on and motivated by the Plasma Liner Experiment (PLX) 
\cite{hsu15jpp,hsu2017experiment}, where six coaxial plasma guns 
\cite{witherspoon17,hsu2017experiment} are mounted on a 2.74-m-diameter spherical vacuum chamber.  
In these experiments, two plasma jets are fired with merging half-angle $\theta=11.6^\circ$ or
20.5$^\circ$, as shown in 
Figs.~\ref{fig:setup}(a) and \ref{fig:setup}(b), respectively. At the gun nozzle, each jet has ion 
density $n_i\sim 2\times 10^{16}$~cm$^{-3}$, electron
temperature $T_e \approx T_i\approx 1.5$~eV, mean-charge
$\bar{Z}\approx 1$, diameter $\approx 8.5$~cm, and speed $v_{\mathrm{jet}}\approx 25$--80~km/s 
\cite{hsu2017experiment}.   Details of the gun 
design and jet characterization are reported elsewhere \cite{witherspoon17,hsu2017experiment}.
Extensive prior work \cite{hsu12pop,merritt2013experimental,merritt14}
showed that a jet propagating over $\sim 1$~m expands radially and axially
at approximately the internal sound speed $C_s$,
$T_e$ and $v_{\mathrm{jet}}$ stay approximately
constant, $n_i$ decreases consistent with mass conservation, magnetic field
strength decays by $1/e$ every few $\mu$s such that both the thermal pressure
and kinetic energy density (of the jet directed motion) dominate over the magnetic pressure 
when the jets merge, and that density jumps and jet-merging morphology are consistent with oblique 
collisional shock formation.  
 
The plasma parameters reported in this work, i.e.,
$T_i$, $T_e$, electron density $n_e$, $\bar Z$, and $v_{\mathrm{jet}}$, are 
inferred from diagnostic measurements (positions shown in Fig.~\ref{fig:setup}).
Plasma $T_i$ is measured via Doppler broadening 
of plasma ion line emission using a high-resolution,
4-m McPherson monochromator (2062DP) with a 2400~mm$^{-1}$ 
grating and an intensified charge-coupled-device (ICCD) detector. The spectral resolution is 
1.5~pm/pixel at the typical visible wavelengths of interest, sufficient to resolve $T_i \gtrsim$ a few eV for 
Xe and correspondingly smaller values for lighter species. 
The high-resolution spectrometer records two chords at a time
with typical waist diameter of 2 cm; chord positions are indicated by the blue dots (10-cm separation) in 
Fig.~\ref{fig:setup}.  Doppler broadening is the 
primary source of line broadening in our parameter regime (the density is too low for
Stark broadening to be appreciable), and the effects of differing Doppler shifts of 
different jets are minimized by viewing the merging at $\approx 90^{\circ}$
relative to the directions of jet propagation. Turbulent motion of the merged plasma is not indicated in the 
experimental images.  Line-integrated measurements of $n_e$ are obtained using a multi-chord laser 
interferometer \cite{merritt2012multi}. The density of the post-shock or jet-interpenetration regions 
are measured using five interferometry chords (0.3-cm chord diameter and 1.5-cm spacing between chords) 30~cm from target-chamber center (TCC), as shown by the green dots in 
Fig.~\ref{fig:setup}. Plasma $T_e$ and $\bar{Z}$ are bounded \cite{hsu12pop}
by comparing broadband visible spectroscopy data \cite{hsu2017experiment},
obtained along the same chord positions to atomic modeling, using the inferred $n_e$ from interferometry. 
Jet speeds are measured via 
a photodiode array at the end of each gun \cite{hsu2017experiment}. 
An ICCD camera (PCO dicam pro) obtains visible-light images
of the shock formation or jet interpenetration.  Further details of the PLX 
facility, coaxial plasma guns, diagnostics, and plasma-jet parameters are described in 
Ref.~\citenum{hsu2017experiment}. 

Figures~\ref{fig:setup}(a) and \ref{fig:setup}(b) show fast-camera images of
two jets merging with $\theta=11.6^\circ$ and 20.5$^\circ$,
respectively, and Fig.~\ref{fig:setup}(c) shows lineouts of
the square root of intensity across the region of jet merging.
If $T_e$ is nearly spatially uniform, which is consistent with both 
collisional plasma-shock theory \cite{jaffrin64} and our experimental measurements,
then the lineouts in Fig.~\ref{fig:setup}(c) are representative
of the $n_i$ profile.  For the black curve, the gradient scale length $\sim$ few cm is
consistent with expected oblique collisional plasma-shock thicknesses (discussed later).

\begin{figure}[!t]
\includegraphics[width=3.25truein]{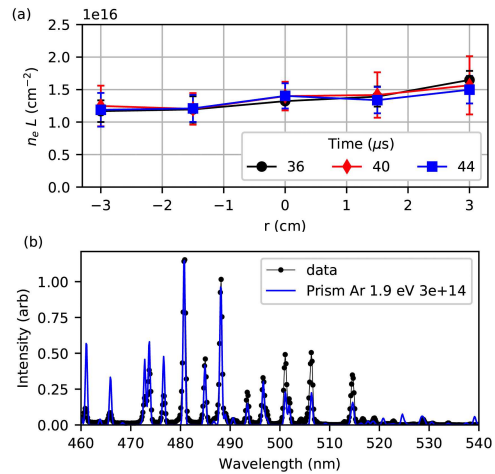}
\caption{(a)~Example line-integrated electron density, at the three indicated times, of 
interpenetrating Ar plasma jets 
(shot 1579, $\theta=20.5^{\circ}$), as measured by interferometry
[green dots in Fig.~\ref{fig:setup}(b)], where $r<0$ is below the midplane and error bars
indicate $\pm 1\sigma$ of shot-to-shot 
variation. (b)~Example visible spectral emission from merged plasma jets
(shot 1579, $t=38$~$\mu$s, 30~cm from TCC)
and calculated spectra using PrismSPECT \cite{macfarlane2003simulation} (uncertainty in
$T_e$ is $\pm 0.4$~eV).
\label{fig:interferom}}
\end{figure}

Figure~\ref{fig:interferom}(a) shows representative interferometry
profiles of line-integrated $n_e$ in the post-merged plasma. These measurements show 
small spatial variations in the post-merge region and are used to infer post-merge
$n_e$.  Figure~\ref{fig:interferom}(b) shows the broadband emission spectrum 
compared to PrismSPECT modeling \cite{macfarlane2003simulation},
which we use to bound $T_e$ and $\bar Z$\@.  
The uncertainties in $T_e$ and $\bar Z$ are 
determined based on the absence/presence of lines compared to 
PrismSPECT modeling \cite{hsu12pop}.  Post-merge values of $n_e$, $T_e$, and $\bar Z$,
are summarized in Table~\ref{tab:conditions}\@. Broadband spectra reveal that no impurity lines 
are observed during the first 10~$\mu$s of jet merging; results in Table~\ref{tab:conditions}
are not expected to be significantly affected by impurities.

\begin{table*}[!t]
\caption{Summary of experimental parameters. The $n_e$, $T_i$, $T_e$, $\bar{Z}$,
and ion--ion mean free path $\lambda_i$ are average, post-merge values.   
The jet--jet interpenetration length $L_{ii,s}$ [see Eq.~(\ref{eqn:ionslowing})],
counter-streaming speed $v=2v_{jet}\sin\theta$,
and jet Mach number $M=v/[\gamma k (T_i + \bar{Z}T_e)/m_i]^{1/2}$
are average, pre-merge values.  The average $L_{ii,s}$ and $\lambda_i$
values are not intended to be precise but to
provide insight into the collisionality regime.  The error ranges for $v_{jet}$, $v$, $n_e$, and $T_i$
are $\pm 1\sigma$ of the variation over multiple shots; those for $T_e$ and $\bar{Z}$
represent uncertainties based on comparisons with PrismSPECT spectral modeling.
\label{tab:conditions}}
\newcolumntype{C}{ @{}>{${}}c<{{}$}@{} }
\begin{ruledtabular}
\begingroup
\setlength{\tabcolsep}{6pt} 
\begin{tabular}{c|*{8}{>{$}r<{$}@{}>{$}l<{$}}>{$}l<{$}}
  Case&\multicolumn{2}{c}{(a)}&\multicolumn{2}{c}{(b)}&\multicolumn{2}{c}{(c)}&\multicolumn{2}{c}{(d)}&\multicolumn{2}{c}{(e)}&\multicolumn{2}{c}{(f)}&\multicolumn{2}{c}{(g)}&\multicolumn{2}{c}{(h)}\\
\hline
  Half-angle $\theta$ &\multicolumn{2}{c}{11.6$^{\circ}$}&\multicolumn{2}{c}{11.6$^{\circ}$}&\multicolumn{2}{c}{11.6$^{\circ}$}&\multicolumn{2}{c}{11.6$^{\circ}$}&\multicolumn{2}{c}{20.5$^{\circ}
  $}&\multicolumn{2}{c}{20.5$^{\circ}$}&\multicolumn{2}{c}{20.5$^{\circ}$}&\multicolumn{2}{c}{20.5$^{\circ}$} \\
  Species&\multicolumn{2}{c}{Ar}&\multicolumn{2}{c}{Xe}&\multicolumn{2}{c}{N}&\multicolumn{2}{c}{Kr}&\multicolumn{2}{c}{Ar}&\multicolumn{2}{c}{Xe}&\multicolumn{2}{c}{N}&\multicolumn{2}{c}{Kr} \\
$v_{\mathrm{jet}}$ (km/s)&41.5&\ \pm\ 4.5&24.3&\ \pm\ 3.1&44.8&\ \pm\ 4.6&64.8&\ \pm\ 18.1&42.1&\ \pm\ 4.8&27.4&\ \pm\ 3.6&52.2&\ \pm\ 3.5&57&\ \pm\ 7.5\\
$v$ (km/s)&16.7&\ \pm\ 1.8&9.8&\ \pm\ 1.2&18.1&\ \pm\ 1.9&26.1&\ \pm\ 7.3&29.4&\ \pm\ 3.3&19.2&\ \pm\ 2.5&36.5&\ \pm\ 2.4&39.8&\ \pm\ 5.3\\
$n_{e}$ ($10^{14}$ cm$^{-3}$)&4.0&\ \pm\ 0.5&4.8&\ \pm\ 0.8&4.6&\ \pm\ 0.4&3.8&\ \pm\ 1.8&4.6&\ \pm\ 1.0&13&\ \pm\ 5.1&8.9&\ \pm\ 1.4&11.6&\ \pm\ 2.9\\
Peak $T_{i}$ (eV)&18.1&\ \pm\ 6.5&25.6&\ \pm\ 3.2&10.2&\ \pm\ 2.2&31.7&\ \pm\ 21.3&32.0&\ \pm\ 2.3&40.6&\ \pm\ 10.0&16.6&\ \pm\ 2.8&45.6&\ \pm\ 10.4\\
  $T_{e}$ (eV)&2.0&\ \pm\ 0.4&1.7&\ \pm\ 0.4&1.7&\ \pm\ 0.9&1.4&\ \pm\ 0.6&2.0&\ \pm\ 0.4&1.7&\ \pm\ 0.4&2.6&\ \pm\ 0.8&1.4&\ \pm\ 0.6\\
$\bar Z$ &1.0&\ \pm\ 0.1&1.2&\ \pm\ 0.2&1.0&\ \pm\ 0.2&1.0&\ \pm\ 0.2&1.0&\ \pm\ 0.1&1.2&\ \pm\ 0.2&1.1&\ \pm\ 0.2&1.0&\ \pm\ 0.2\\
$L_{ii,s}$ (cm)&\multicolumn{2}{c}{2.5}&\multicolumn{2}{c}{1.5}&\multicolumn{2}{c}{0.2}&\multicolumn{2}{c}{56.2}&\multicolumn{2}{c}{26.6}&\multicolumn{2}{c}{10.2}&\multicolumn{2}{c}{4.0}&\multicolumn{2}{c}{190} \\
$\lambda_{i}$ (cm)&\multicolumn{2}{c}{1.9}&\multicolumn{2}{c}{1.6}&\multicolumn{2}{c}{0.5}&\multicolumn{2}{c}{2}&\multicolumn{2}{c}{3.3}&\multicolumn{2}{c}{1.4}&\multicolumn{2}{c}{0.4}&\multicolumn{2}{c}{2.6} \\
$M$&\multicolumn{2}{c}{4.2}&\multicolumn{2}{c}{4.7}&\multicolumn{2}{c}{2.9}&\multicolumn{2}{c}{11.4}&\multicolumn{2}{c}{7.4}&\multicolumn{2}{c}{9.1}&\multicolumn{2}{c}{4.6}&\multicolumn{2}{c}{17.3} \\
\end{tabular}
\endgroup
\end{ruledtabular}
\end{table*}

\begin{figure}[!t]
\includegraphics[width=3.25truein]{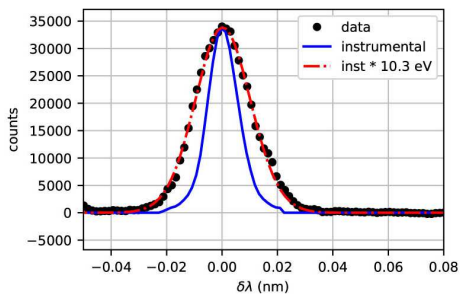}
\caption{Example of high-resolution spectroscopy data and fitting to infer $T_i=10.3$~eV
based on the best fit of the convolution of instrumental broadening with
a Gaussian function (shot 1601, $t=32$~$\mu$s, 1-$\mu$s gate,
$\theta=11.6^\circ$, Ar~{\sc ii} 480.6-nm line,
30~cm from TCC, fitting error $=\pm 0.3$~eV)\@.\label{fig:doppler}}
\end{figure}

The primary result of this work is the measurement of the time evolution of $T_i$
inferred from Doppler broadening of ionized emission lines, in the post-shock plasma
or the region of jet--jet interpenetration as shown in Figs.~\ref{fig:setup}(a) and \ref{fig:setup}(b),
respectively.  An example of the inference of $T_i$ from Doppler spectroscopy data is shown in 
Fig.~\ref{fig:doppler}.  Data at the earliest stage of jet merging show evidence of multiple 
overlapping line shapes (not shown here), which we believe to be due to interpenetration and
systematic gun-angle-dependent Doppler shifts. These features are not observed several $\mu$s
later into the jet merging. In data processing, we reject multiple-line-shape cases
and include only the cases that satisfy a threshold goodness-of-fit to a single Gaussian.

\begin{figure}[!htb]
\includegraphics[width=3.3truein]{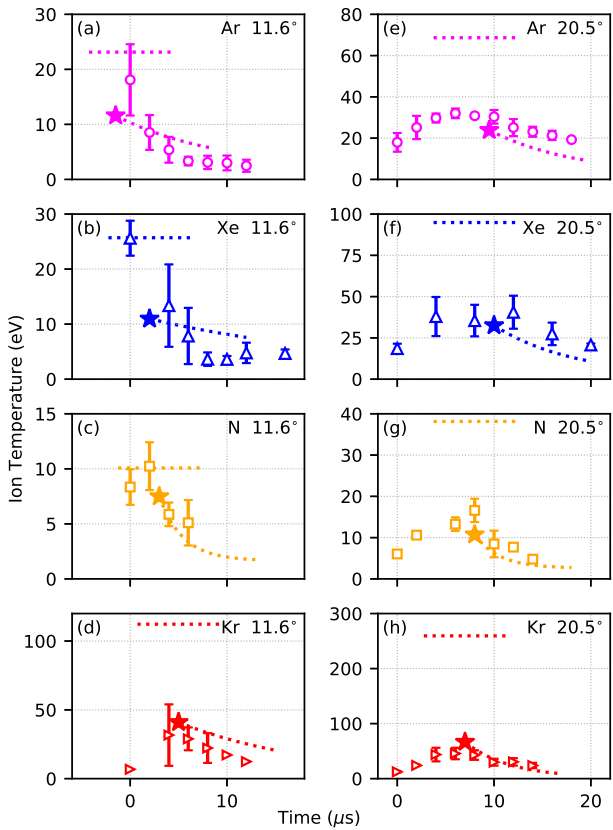}
\caption{Measured $T_i$, inferred from Doppler spectroscopy, vs.\ time
corresponding to cases (a)--(h) of Table~\ref{tab:conditions}
[shot ranges 1594--1625, 1744--1776, 
2606--2619, 2307--2346, 1563--1593, 1717--1743, 2139--2168, and 2271--2306, respectively].
Error bars indicate $\pm 1\sigma$ variation 
across $\approx 5$ shots per data point (where available). Horizontal dotted lines denote peak $T_i$  
based on Eq.~(\ref{eqn:shockiontemp}).  Dotted
lines overlaying the data are ion--electron temperature relaxation
based on Eq.~(\ref{eqn:relax}). Stars indicate peak 
$T_i$ from 1D-equivalent multi-fluid simulations (see text; star positions are
not intended to reflect the simulation time). 
The $T_e\lesssim 3$~eV in all cases (see Table~\ref{tab:conditions}).\label{fig:ion_temp}}
\end{figure}

Figure~\ref{fig:ion_temp} shows inferred $T_i$ versus time corresponding to cases (a)--(h) of 
Table~\ref{tab:conditions}\@.
Cases (a)--(c) and (g) are expected to be ``collisional'' with oblique shock formation [e.g.,
Fig.~\ref{fig:setup}(a)], while cases (d)--(f) and (h) are expected to be ``interpenetrating'' without
apparent shock formation [e.g., Fig.~\ref{fig:setup}(b)].
``Collisional'' and ``interpenetrating'' are defined in the next paragraph.
Specific emission lines used were 463.0-nm N~{\sc ii}, 
480.6-nm Ar~{\sc ii}, 473.9-nm Kr~{\sc ii}, and 529.2-nm Xe~{\sc ii}\@.
In obtaining this dataset at the positions
indicated by the blue dots in Figs.~\ref{fig:setup}(a) and \ref{fig:setup}(b), we recorded progressively
later times as we moved the spectrometer viewing chords closer to TCC (over multiple shots)
because the jets and merged plasma move from right to left in Figs.~\ref{fig:setup}(a) 
and \ref{fig:setup}(b).
All recorded data meeting the goodness-of-fit criterion are included in Fig.~\ref{fig:ion_temp}.

We consider the approximate interpenetration distance $L_{ii,s}$
between merging jets, which can vary from much smaller (``collisional'') to of the order or greater 
(``interpenetrating'')
than the characteristic jet size $L\sim 20$~cm.  Using average pre-merge jet parameters
($v_{\mathrm{jet}}$ from photodiodes, $n_i$ decreased from measured 
post-merge $n_i=n_e/\bar Z$ by a factor of 2.5 for interpenetrating cases and 3.5 for 
shock-forming cases, which are
approximations between theoretical limits of 2 and 4, respectively, for $\gamma = 
5/3$, and $\bar{Z}$ inferred from 
spectroscopy), we estimate \cite{formulary16} 
\begin{equation}
L_{ii,s} = \frac{v}{4 \nu_{ii,s}} = \frac{v}{4} \left[\num{9e-8}n_i \bar Z^4\Lambda_{ii}  \left( \frac{2}{\mu}  \right) 
\frac{\mu^{1/2}}{\epsilon^{3/2}} \right]^{-1},
\label{eqn:ionslowing}
\end{equation}
where $v=2v_{\mathrm{jet}}\sin\theta$ (cm/s) is the counter-streaming speed between the two jets,
$\nu_{ii,s}$ the counter-streaming ion--ion slowing frequency in the fast limit
($\gg \nu_{ie,s}$ for our parameters),
$\Lambda_{ii}$ the Coulomb logarithm for counter-streaming ions in the presence
of warm electrons \cite{merritt14,formulary16}, $\mu$ the ion/proton mass ratio,
$\epsilon$ (eV) the energy associated with $v$, and the factor of 4 in the denominator accounts for the 
integral effect of slowing down
\cite{messer13pop}. For cases (a)--(c) and (g) of Table~\ref{tab:conditions},
$L \gg L_{ii,s}$.  For cases (d)--(f) and (h) of Table~\ref{tab:conditions},
$L \lesssim L_{ii,s}\sim v \epsilon^{3/2} \sim v^4$.

If $L_{ii,s}\ll L$, the jets impact each other like pistons, and collisional
plasma shocks typically form \cite{merritt2013experimental,merritt14}.
An upper bound for the jump in $T_i$ across the shock, assuming that all of the heating goes to the
ions, and $T_e$ is uniform across the shock, is \cite{suppPRL,liepmann1957elements}
\begin{equation}
\frac{T_{i2}}{T_{i1}}  = \left[ 1 + \frac{2 (\gamma - 1)}{(\gamma + 1)^2} \frac{\gamma M^2 + 1}{M^2} (M^2 - 1) \right] \left( \alpha + 1 \right) - \alpha,
\label{eqn:shockiontemp}
\end{equation}
where subscripts `1' and `2' refer to pre- and post-shock, respectively, 
$\gamma=5/3$ is the polytropic index, pre-shock Mach number
$M \equiv v / [\gamma k (T_i + \bar Z T_e)/m_i]^{1/2}$, $\alpha \equiv (\bar Z T_e) / T_{i1}$,
and $T_{i1}=T_e$ is assumed. Predicted $T_{i2}$ based on 
Eq.~(\ref{eqn:shockiontemp}) are plotted as horizontal dotted lines in Fig.~\ref{fig:ion_temp}.
Figure~\ref{fig:ion_temp} shows that the measured peak $T_i$ agrees well with
Eq.~(\ref{eqn:shockiontemp}) for shock-forming cases with $L_{ii,s} \ll L$ [Figs.~\ref{fig:ion_temp}(a)--(c)], 
becomes increasingly smaller than predicted with increasing $L_{ii,s}$
[Figs.~\ref{fig:ion_temp}(a) and (g)], and is uniformly much smaller than predicted when
$L_{ii,s}\gtrsim L$ [Figs.~\ref{fig:ion_temp}(d)--(f) and (h)].
Within each species, the measured $T_i$ evolution becomes less impulsive and has a broader 
temporal profile with increasing $L_{ii,s}$.
When $L_{iis,s}\ll L$ [e.g., Fig.~\ref{fig:setup}(a)], 
the estimated post-shock ion--ion mean free paths $\lambda_i \sim 1$~cm, 
consistent with the sharp jumps of the solid black curve of Fig.~\ref{fig:setup}(c) being 
collisional shocks.  When $L_{ii,s}\gtrsim L$,
shocks do not appear to form [e.g., Fig.~\ref{fig:setup}(b) and
blue dotted curve of Fig.~\ref{fig:setup}(c)].

The predicted, classical ion--electron temperature relaxation rate \cite{formulary16},
\begin{equation}
\frac{dT_i}{dt} = \left[ \num{1.8e-19} \frac{ \left(m_i m_e \right)^{1/2}  \bar{Z}_i^2 n_e \Lambda_{ie} }
{ \left( m_i T_e + m_e T_i \right)^{3/2}} \right] \left( T_e - T_i \right),
\label{eqn:relax}
\end{equation}
is plotted in Fig.~\ref{fig:ion_temp}, overlaying the data.
Agreement between the data and Eq.~(\ref{eqn:relax})
is generally good.  Discrepancies beyond the error bars
motivate further detailed comparisons with theory/modeling, e.g., accounting for 
multi-dimensional, radiative, and EOS effects.

Finally, we perform 1D (counter-streaming component),
multi-fluid calculations (Lagrangian particles advect electron- and two ion-fluid 
quantities), including thermal/radiative losses and tabular EOS, of peak $T_i$ (stars in
Fig.~\ref{fig:ion_temp}) using the Chicago code \cite{thoma11pop, thoma2017hybrid}. 
For collisional cases [Figs.~\ref{fig:ion_temp}(a)--(c) and (g)],
calculated peak $T_i$ are lower than Eq.~(2) (expected with 
inclusion of thermal/radiative losses) but are also somewhat lower than the data.  For interpenetrating
cases [Figs.~\ref{fig:ion_temp}(d)--(f) and (h)], the calculated peak $T_i$ agree reasonably well with
the data. Remaining discrepancies motivate detailed, multi-dimensional validation studies beyond
the scope of this work.

In conclusion, we report a comprehensive
experimental study of ion heating in collisional plasma shocks 
and interpenetrating supersonic plasma flows formed by 
the oblique merging of two laboratory plasma jets.  The post-merge $T_i \gg T_e$ in all cases investigated, 
including for
both very small and substantial jet interpenetration, indicating that the predominant heating goes
to the ions for both cases. For cases with
shock formation, the measured peak $T_i$ agrees in most cases with the theoretically predicted $T_i$
jump for a collisional plasma shock [Eq.~(\ref{eqn:shockiontemp})].
For interpenetrating cases, the measured peak $T_i$, unsurprisingly, is substantially below
that predicted by collisional plasma-shock theory.  The predicted classical ion--electron temperature 
relaxation compares reasonably well with the observed $T_i$ decay.  
Multi-fluid Chicago simulations show some agreement with the peak-$T_i$
data in both shock-forming and interpenetrating cases; the differences highlight an 
opportunity for detailed, multi-dimensional model 
validation for this and other codes being used to design and advance
our understanding of HED and ICF experiments.

\begin{acknowledgments}
We acknowledge J. Dunn, E. Cruz, A. Case, F. D. Witherspoon, S. Brockington, 
J. Cassibry, R. Samulyak, P. Stoltz, Y. C. F. Thio, and D. Welch
for technical support and/or useful discussions. This work was supported by the Office of Fusion Energy 
Sciences and the Advanced Research Projects Agency--Energy of the U.S. Dept.\ of Energy under 
contract no.\ DE-AC52-06NA25396.
\end{acknowledgments}

\providecommand{\noopsort}[1]{}\providecommand{\singleletter}[1]{#1}%

\end{document}